\documentclass[aps,prb,twocolumn,superscriptaddress]{revtex4-1}

\usepackage{amsmath}
\usepackage[margin = 50pt]{geometry}
\usepackage{graphicx}
\usepackage{epstopdf}
\usepackage{dcolumn}
\usepackage[caption=false]{subfig}
\usepackage{float}

\begin{document}


\title{Decomposition of scattered electromagnetic fields into vector spherical wave functions on surfaces with general shapes}


\author{X. Garcia Santiago}
\email[]{xavier.garcia-santiago@kit.edu}
\affiliation{Institute of Nanotechnology, Karlsruhe Institute of Technology, 76021 Karlsruhe, Germany}
\affiliation{JCMwave GmbH, 14050 Berlin, Germany}
\author{M. Hammerschmidt}
\affiliation{JCMwave GmbH, 14050 Berlin, Germany}
\author{S. Burger}
\affiliation{JCMwave GmbH, 14050 Berlin, Germany}
\affiliation{Zuse Institute Berlin, 14195 Berlin, Germany}
\author{C. Rockstuhl}
\affiliation{Institute of Nanotechnology, Karlsruhe Institute of Technology, 76021 Karlsruhe, Germany}
\affiliation{Institute of Theoretical Solid State Physics, Karlsruhe Institute of Technology, 76131 Karlsruhe, Germany}
\author{I. Fernandez-Corbaton}
\affiliation{Institute of Nanotechnology, Karlsruhe Institute of Technology, 76021 Karlsruhe, Germany}
\author{L. Zschiedrich}
\affiliation{JCMwave GmbH, 14050 Berlin, Germany}

\date{\today}

\begin{abstract}
Decomposing the field scattered by an object into vector spherical wave functions (VSWF) is a useful tool when discussing its optical properties on more analytical grounds. Thus far, it was frequently required in the decomposition that the scattered field is available on a spherical surface enclosing the scatterer. This requirement is adapted to the spatial dependency of the VSWFs but is rather incompatible with many numerical solvers. To mitigate this problem, we propose an orthogonal expression for the decomposition that holds for any surface that encloses the scatterer, independently of its shape. 
We also show that the orthogonal relations remain unchanged when the radiative VSWF used for the expansion of the scattered field are substituted by the regular VSWF used for the expansion of the incident illumination as test functions. This is a key factor for the numerical stability of our decomposition. As example, we use a finite-element based solver to compute the multipole response of a nanorod illuminated by a plane wave and study its convergence properties.
\end{abstract}


\maketitle

\section{Introduction}

The vector spherical wave function (VSWF) decomposition, or multipole expansion, is a useful tool to study electromagnetic scattering phenomena. Vector spherical wave functions are well known solutions to the time harmonic Maxwell's equations in homogeneous media \cite{jackson1999classical, mishchenko2002scattering}, forming a complete basis for the electromagnetic field. The field scattered by an object immersed in a homogeneous medium upon interacting with an incident field can be decomposed into the radiative VSWFs $\mathbf{N}_{m,n}^{(3)}\left(\mathbf{r}\right)$ and $\mathbf{M}_{m,n}^{(3)}\left(\mathbf{r}\right)$ as,

\begin{equation}\label{eq:E_scat_expansion}
\mathbf{E_{scat}}(\mathbf{r}) = \sum_{n = 1}^{\infty}{\sum_{m = -n}^{n}{\left[a_{m,n}\mathbf{N}_{m,n}^{(3)}\left( \mathbf{r}\right)+b_{m,n}\mathbf{M}_{m,n}^{(3)}\left( \mathbf{r}\right)\right]}}.
\end{equation}
This decomposition is only valid in the region outside the smallest sphere circumscribing the scatterer object \cite{RH_Auguie,RH_Bates,RH_Berg}. The elements of the basis correspond to the field created by a point multipole with definite properties: $n$ refers to the total angular momentum and $m$ to the angular momentum along a chosen axis. $\mathbf{N}_{m,n}^{(3)}\left( \mathbf{r}\right)$ and $\mathbf{M}_{m,n}^{(3)}\left( \mathbf{r}\right)$ are multipolar fields with different parity symmetry and their definition is shown in Appendix \ref{sec:Appendix_VSWF}. They correspond to the electric field of electric multipoles and the electric field of magnetic multipoles, respectively. The complex amplitudes $a_{m,n}$ and $b_{m,n}$ in the expansion express then the contribution of the respective multipolar field to the total scattered field. It is the purpose of the multipole expansion to identify these amplitudes.

These amplitudes contain valuable information about the interaction of light with the scatterer. In consequence, the decomposition is used in many streams of research. Prime examples would be the study of optical nanoantennas~\cite{Coenen2013,Rusak2014,stout2011multipole, dielectric_nanoantennas, Pors:10}, the study of meta-atoms and metamaterials~\cite{Liu2017, grahn2012electromagnetic, Powell2017, petschulat2010understanding} or the analysis of the interaction of scatterers with isolated molecules~\cite{Vercruysse2014,Hongming2016,Heaps2017}. 
Using the multipole expansion we can also construct the T-matrix of a scatterer that entirely expresses how an arbitrary incident field is scattered by the pertinent  object~\cite{T_matrix,demesy2018scattering}. Once the T-matrix of different individual scatterers is known, the interaction of light with a larger cluster of heterogeneous particles can be solved using a multiple-scattering algorithm~\cite{Light_scattering_systems_of_particles,GIMBUTAS_Fast_multi_particle_scattering}. 

For spherical particles, the multipole expansions can be calculated analytically using Mie theory \cite{wriedt_mie}. However, for more complex structures, the use of numerical solvers is needed in order to obtain the scattered field first and to decompose it afterwards \cite{Martin_Ivan_T}. Among other methods, FEM is specially suitable for the first task, as it can accurately deal with structures possessing complex shapes. When the scattered field ($\mathbf{E_{scat}}\left(\mathbf{r}\right)$) produced by a particle under a certain illumination has been calculated, the multipole coefficients can be obtained thanks to the orthogonality relations of the VSWFs

\begin{eqnarray}
\frac{\int_{S^2_R}{\mathbf{N}_{m1,n1}^{(J)*}\left(\mathbf{r}\right)\cdot\mathbf{N}_{m2,n2}^{(J)}\left(\mathbf{r}\right) \, \mathrm{dS}}}{\int_{S^2_R}{|\mathbf{N}_{m1,n1}^{(J)}\left(\mathbf{r}\right)|^2 \,\mathrm{dS}}} &=& \delta _{m1m2} \delta _{n1n2}, \label{eq:orthogonal1}\\
\frac{\int_{S^2_R}{\mathbf{M}_{m1,n1}^{(J)*}\left(\mathbf{r}\right)\cdot\mathbf{M}_{m2,n2}^{(J)}\left(\mathbf{r}\right) \, \mathrm{dS}}}{\int_{S^2_R}{|\mathbf{M}_{m1,n1}^{(J)}\left(\mathbf{r}\right)|^2 \,\mathrm{dS}}} &=& \delta _{m1m2} \delta _{n1n2}, \label{eq:orthogonal2}\\
\int_{S^2_R}{\mathbf{M}_{m1,n1}^{(J)*}\left(\mathbf{r}\right)\cdot\mathbf{N}_{m2,n2}^{(J)}}\left(\mathbf{r}\right) \, \mathrm{dS} &=& 0, \label{eq:orthogonal3}
\end{eqnarray}
where the integrals are on the surface $S^2_R$ of a sphere of radius R centered at the origin of coordinates. The above expressions hold for the radiative VSWFs (J=3), which fulfill the radiation condition and can be used for decomposing a scattered field, and for the regular VSWFs (J=1), used for example for expressing an illumination field in terms of multipoles. 

Therefore, coefficients $a_{m,n}$ and $b_{m,n}$ in Eqn.~(\ref{eq:E_scat_expansion}) can be obtained by computing the integrals

\begin{align}
a_{m,n} = \frac{\int_{S^2_R}{\mathbf{N}_{m,n}^{(3)*}\left(\mathbf{r}\right)\cdot\mathbf{E_{scat}}\left(\mathbf{r}\right) \,\mathrm{dS}}}{\int_{S^2_R}{|\mathbf{N}_{m,n}^{(3)}\left(\mathbf{r}\right)|^2} \,\mathrm{dS}}, \label{eq:a_decomposition}\\
b_{m,n} = \frac{\int_{S^2_R}{\mathbf{M}_{m,n}^{(3)*}\left(\mathbf{r}\right)\cdot\mathbf{E_{scat}}\left(\mathbf{r}\right) \,\mathrm{dS}}}{\int_{S^2_R}{|\mathbf{M}_{m,n}^{(3)}\left(\mathbf{r}\right)|^2} \,\mathrm{dS}}. \label{eq:b_decomposition}
\end{align}

However, the fact that the domain of integration has to be a perfect sphere has some drawbacks. First, the numerical solution $\mathbf{E_{scat}}\left(\mathbf{r}\right)$ must be interpolated across the sphere in order to perform the decompositions ~(\ref{eq:a_decomposition})-(\ref{eq:b_decomposition}). This interpolation produces accuracy losses and it is also computationally expensive, considerably increasing the total computation time. 

One possible solution to overcome these drawbacks is to perform the expansion based on volume integrals of the currents induced in the scatterer structures~\cite{grahn2012electromagnetic,FerCor2015b,Alaee2016b}. In this work, we propose a different approach, using the orthogonality property for VSWF extended to any closed surface. In this way, any surface which encloses the scatterer can be used to perform the decomposition, including the boundary of the computational domain or the surface of the scatterer. It is normally easier to implement fast and highly-accurate surface integration methods over these natural boundaries of the problem. Furthermore, most of the available FEM and FDTD software have already efficient built-in implementations for these integrals that we can profit from.

\section{Decomposition of the scattered field over a general surface}

We derive in the following an expression to decompose the scattered field into VSWFs via an integration over a surface with a non-constrained shape. We assume that the scatterer is localized in space and surrounded by a homogeneous, isotropic, and lossless medium.

Let us denote by $\mathbf{E_{scat}}\left(\mathbf{r}\right)$ the field solution of our time-harmonic scattering problem and by $\mathbf{F}\left(\mathbf{r}\right)$ another field solution of Maxwell's equations. Under the above conditions, Maxwell's equations can be written as 

\begin{eqnarray}
\nabla\times\nabla\times\mathbf{E_{scat}}\left(\mathbf{r}\right) -k^2\mathbf{E_{scat}}\left(\mathbf{r}\right) &=& 0, \label{eq:Maxwell_E}\\
\nabla\times\nabla\times\mathbf{F}\left(\mathbf{r}\right) -k^2\mathbf{F}\left(\mathbf{r}\right) &=& 0, \label{eq:Maxwell_F}
\end{eqnarray}
with $k$ being the wave number $k = \omega \sqrt{\mu \epsilon}$, $\mu$ and $\epsilon$ are the magnetic permeability and the electric permittivity of the homogeneous medium, respectively, and $\omega$ is the angular frequency of the field.

We consider that both fields fulfil an outwards Silver-M\"uller radiation condition \cite{kirsch2016mathematical} 

\begin{align}\label{eq:Silver-Mueller}
&\lim_{r\to\infty}\left[ \left(\frac{1}{\sqrt{\mu}}\nabla \! \times \! \left\{\mathbf{E_{scat}},\mathbf{F}\right\}\left(\mathbf{r}\right)\right) \! \times \! \mathbf{r} 
\! - \! |\mathbf{r}|i\omega\sqrt{\epsilon}\left\{\mathbf{E_{scat}},\mathbf{F}\right\}\left(\mathbf{r}\right) \right]  \\ &= 0. \nonumber
\end{align}

We want to obtain a scalar product, which shows an orthogonality relation similar to expressions~(\ref{eq:orthogonal1}) to (\ref{eq:orthogonal3}), but featuring an integral over a surface with a shape not restricted to a sphere. In order to obtain this expression, we start by multiplying Eqn.~(\ref{eq:Maxwell_E}) with the complex conjugate of the field $\mathbf{F}$

\begin{equation}\label{eq:E_dot_F}
\mathbf{F^*}\cdot\left\{\nabla\times\left(\nabla\times\mathbf{E_{scat}}\right)\right\} -k^2\mathbf{F^*}\cdot\mathbf{E_{scat}} = 0. 
\end{equation}
In this derivation we omit for brevity the explicit position dependency $\left(\mathbf{r}\right)$ of the fields $\mathbf{E_{scat}}$, $\mathbf{F}$, $\mathbf{M}_{m,n}^{(J)}$ and $\mathbf{N}_{m,n}^{(J)}$ but it is implicitly assumed all the time.

\begin{figure}[htpb]
\centering
\includegraphics[width=0.6\linewidth]{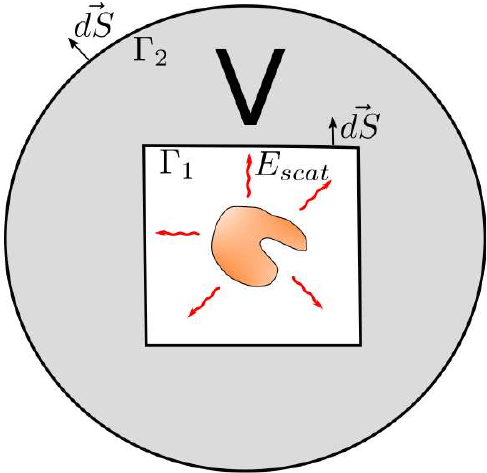}
\caption{Illustration of the problem considered in this contribution. An object of arbitrary shape generates some scattered field everywhere in the outer region. We consider the integration of Eqn.~(\ref{eq:E_dot_F}) in a volume $V$ surrounding the scattered object. The volume is delimited by two surface boundaries $\Gamma _1$ and $\Gamma _2$. Eventually, only the integral across $\Gamma _1$ is used to express the multipole coefficients. The shape of this surface can be arbitrary.}
\label{fig:esquema_integral}
\end{figure}

Integrating Eqn.~(\ref{eq:E_dot_F}) over a volumetric region $V$ that surrounds but does not contain any part of the scatterer (see Fig.~\ref{fig:esquema_integral}) we get
\begin{equation}
\int_V{\left(\mathbf{F^*}\cdot\left\{\nabla\times\left(\nabla\times\mathbf{E_{scat}}\right)\right\} -k^2\mathbf{F^*}\cdot\mathbf{E_{scat}} \right)\,dV} = 0. 
\end{equation}
Applying two partial integrations over the integration volume 

\begin{equation}
\begin{split}
0 =&\int_{\Gamma}{\left\{\left(\nabla\times\mathbf{E_{scat}}\right)\times \mathbf{F^*}\right\}\cdot\mathbf{dS}} \\
+&\int_V{\big\{\left(\nabla\times\mathbf{E_{scat}}\right) \cdot \left(\nabla\times\mathbf{F^*}\right) -k^2\mathbf{F^*}\cdot\mathbf{E_{scat}} \big\} dV} \\
=&\int_\Gamma{\left\{\left(\nabla\times\mathbf{E_{scat}}\right)\times \mathbf{F^*} + \mathbf{E_{scat}}\times\left(\nabla\times\mathbf{F^*}\right) \right\}\cdot\mathbf{dS}} \\
+&\int_V{\left(\mathbf{E_{scat}}\cdot\left\{\nabla\times\left(\nabla\times\mathbf{F^*}\right)\right\} -k^2\mathbf{F^*}\cdot\mathbf{E_{scat}} \right) dV} \\
\end{split}
\end{equation}
and exploiting the fact that $\mathbf{F}$ fulfils Maxwell's equations (Eqn.~(\ref{eq:Maxwell_F}))

\begin{equation}
\begin{split}
&\int_V{\left(\mathbf{E_{scat}}\cdot\left\{\nabla\times\left(\nabla\times\mathbf{F^*}\right)\right\} -k^2\mathbf{F^*}\cdot\mathbf{E_{scat}} \right) dV} \\
=&\int_V{\left(\left\{\nabla\times\left(\nabla\times\mathbf{F^*}\right) -k^2\mathbf{F^*}\right\}\cdot\mathbf{E_{scat}} \right) dV} = 0,
\end{split}
\end{equation}
we get

\begin{equation}
\int_{\Gamma}{\left\{\left(\nabla\times\mathbf{E_{scat}}\right)\times\mathbf{F^*}-\left(\nabla\times\mathbf{F^*}\right)\times\mathbf{E_{scat}}\right\}\cdot\mathbf{dS}} = 0, 
\end{equation}
where $\Gamma$ is the boundary of the volume V.

Now we split the surface integral into the contribution of an inner surface $\Gamma _1$ and an outer surface $\Gamma _2$ (see Fig.~\ref{fig:esquema_integral}),

\begin{align}\label{eq:surface_integrals_F}
- &\int_{\Gamma _1}{\left\{\left(\nabla\times\mathbf{E_{scat}}\right)\times\mathbf{F^*}-\left(\nabla\times\mathbf{F^*}\right)\times\mathbf{E_{scat}}\right\}\cdot\mathbf{dS}} \nonumber \\
+ &\int_{\Gamma _2}{\left\{\left(\nabla\times\mathbf{E_{scat}}\right)\times\mathbf{F^*}-\left(\nabla\times\mathbf{F^*}\right)\times\mathbf{E_{scat}}\right\}\cdot\mathbf{dS}} \nonumber \\ &= I_1+I_2 = 0. 
\end{align}
Integrals in the previous equation are defined to have $\mathbf{dS}$ elements pointing inwards and outwards the volume $V$ in $\Gamma _1$ and $\Gamma _2$, respectively, as seen in Fig.~\ref{fig:esquema_integral}. 

As we have not imposed any condition on $\Gamma_1$ or $\Gamma_2$, the integral values must hold independently of the shape of the surfaces. Since we want an expression that equals relations~(\ref{eq:orthogonal1})-(\ref{eq:orthogonal3}), we will consider in the following the exterior surface $\Gamma_2$ to be a sphere of radius R ($S^2_R$). 

Now we will apply the circular shift invariance of the scalar triple product to the terms of $I_2$. The reason for this change will be clear a few lines below. It reads as

\begin{align}\label{eq:I2}
I_2 &\!=\! \int_{S^2_R}{\left\{\left(\nabla \! \times \! \mathbf{E_{scat}}\right) \! \times \! \mathbf{F^*}\right\}\!\cdot\!\mathbf{dS}\!-\!\left\{\left(\nabla \! \times \! \mathbf{F^*}\right) \!\times \! \mathbf{E_{scat}}\right\} \!\cdot \! \mathbf{dS}} \nonumber \\
&\!= \! \int_{S^2_R}{\! \left\{\mathbf{dS} \! \times \! \left(\nabla \! \times \! \mathbf{E_{scat}}\right)\right\} \! \cdot \!\mathbf{F^*} \! - \! \left\{\mathbf{dS} \! \times \! \left(\nabla \! \times \! \mathbf{F^*}\right)\right\} \! \cdot \! \mathbf{E_{scat}}}.
\end{align}

Since $I_2=-I_1$, $I_2$ must be independent of the radius $R$ of the sphere. We can assume it to be big enough for the far-field approximation to hold. Using spherical coordinates, the Silver-M\"uller radiation boundary condition given in Eqn.~(\ref{eq:Silver-Mueller}) reads

\begin{equation}
\lim_{R\to\infty}{\left(\nabla\times\left\{\mathbf{E_{scat}},\mathbf{F}\right\}\right)\times R\mathbf{\hat{r}}} =  +\lim_{R\to\infty}{ikR\left\{\mathbf{E_{scat}},\mathbf{F}\right\}}.
\end{equation}

By expressing the differential surface element of a sphere in spherical coordinates,
\begin{equation}
\mathbf{dS} = R^2\sin{\theta}\, \mathrm{d\theta} \, \mathrm{d\phi} \, \mathbf{\hat{r}},
\end{equation}
we can formulate the Silver-M\"uller radiation condition as

\begin{equation}\label{eq:Silver-Mueller_with_dS}
\lim_{R\to\infty}{\left(\nabla\times\left\{\mathbf{E_{scat}},\mathbf{F}\right\}\right)\times \mathbf{dS}} = + \lim_{R\to\infty}{ik\mathrm{dS}\,\left\{\mathbf{E_{scat}},\mathbf{F}\right\}}.
\end{equation}

Both fields $\mathbf{E_{scat}}$ and $\mathbf{F}$ fulfil the outward radiation boundary condition Eqn.~(\ref{eq:Silver-Mueller}). $\mathbf{F^*}$ fulfils the inward radiation boundary condition, which is obtained changing the minus sign for a plus sign in Eqn.~(\ref{eq:Silver-Mueller}). We use Eqn.~(\ref{eq:Silver-Mueller_with_dS}) in Eqn.~(\ref{eq:I2}) to get

\begin{align} \label{eq:final_I2}
\lim_{R\to\infty} I_2 = -\lim_{R\to\infty} ik\int_{S^2_R}{\left(\mathbf{E_{scat}}\cdot\mathbf{F}^*\ + \mathbf{F}^*\cdot\mathbf{E_{scat}}\right)\,\mathrm{dS}}.
\end{align}
Note that the last expression holds only in the far-field. 
 
Finally, substituting $\mathbf{F}$ by $\mathbf{M}_{m,n}^{(3)}$ or $\mathbf{N}_{m,n}^{(3)}$ into Eqn.~(\ref{eq:final_I2}), we get from Eqns.~(\ref{eq:surface_integrals_F}) and (\ref{eq:final_I2})

\begin{equation}\label{eq:J3_far_field_identity}
\begin{split}
&\int_{\Gamma _1}\bigg\{\left(\nabla\times\mathbf{E_{scat}}\right)\times\left\{\mathbf{M},\mathbf{N}\right\}_{m,n}^{(3)*}\\
 &-\left(\nabla\times\left\{\mathbf{M},\mathbf{N}\right\}_{m,n}^{(3)*}\right)\times\mathbf{E_{scat}}\bigg\}\cdot\mathbf{dS} \\
&=\lim_{R\to\infty} 2ik\int_{S^2_R}{\mathbf{E_{scat}}\cdot\left\{\mathbf{M},\mathbf{N}\right\}_{m,n}^{(3)*}\,\mathrm{dS}}. 
\end{split}
\end{equation}
In comparison with Eqns.~(\ref{eq:a_decomposition}) and (\ref{eq:b_decomposition}), we see that the last integral provides the coefficient of the decomposition of the scattered field into the basis of VSWF times the norm of the VSWF in the far-field, i.e.

\begin{equation}\label{eq:result}
\begin{split}
&2ik \lim_{R\to\infty}\int_{S^2_R}{\mathbf{E_{scat}}\cdot\left\{\mathbf{M},\mathbf{N}\right\}_{m,n}^{(3)*}\,\mathrm{dS}} \\
&= \left\{a,b\right\}_{m,n}2ik\lim_{R\to\infty}\int_{S^2_R}{|\left\{\mathbf{M},\mathbf{N}\right\}_{m,n}^{(3)}|^2 \,\mathrm{dS}}\\
&= \left\{a,b\right\}_{m,n}\frac{2i}{k},
\end{split}
\end{equation}
where we have used that the integral of the norm of every VSWF has a value of $1/k^2$ in the far-field (cf. Ref.~\onlinecite{mishchenko2002scattering}, Eqn.~(C.152)).
We, therefore, conclude from Eqns. (\ref{eq:J3_far_field_identity}) and (\ref{eq:result}) that we can decompose the scattered field into the vector spherical wave functions as the integral over the boundary $\Gamma _1$

\begin{equation}\label{eq:final_with_J3}
\begin{split}
\left\{a,b\right\}_{m,n} = \frac{k}{2i}\int_{\Gamma _1}&{\bigg\{  \left(\nabla\times\mathbf{E_{scat}}\right)\times\left\{\mathbf{M},\mathbf{N}\right\}_{m,n}^{(3)*}} \\
&-k\left\{\mathbf{N},\mathbf{M}\right\}_{m,n}^{(3)*}\times\mathbf{E_{scat}}\bigg\}\cdot\mathbf{dS},
\end{split}
\end{equation}
where we have used the well known properties \cite{mishchenko2002scattering} $\nabla\times\mathbf{M}=k\mathbf{N}$ and $\nabla\times\mathbf{N}=k\mathbf{M}$ to replace $\nabla\times\left\{\mathbf{M},\mathbf{N}\right\}_{m,n}^{(3)}$ by $k\left\{\mathbf{N},\mathbf{M}\right\}_{m,n}^{(3)}$.

We note that it is also possible to reach Eqn.~(\ref{eq:final_with_J3}) by starting from Lemma~6.38 of Ref.~\onlinecite{colton2012inverse} and applying it to $\mathbf{E_{scat}}$ and the VSWF basis functions. 

Expression~(\ref{eq:final_with_J3}) lets us implement the multipole expansion in an easier way, without the need to perform any expensive interpolation to a spherical domain. This is done by including an explicit contribution of the rotational of the electric field into the orthogonality expression, in contrast with Eqns.~(\ref{eq:a_decomposition})-(\ref{eq:b_decomposition}).

Note that even if the expansion in Eqn.~(\ref{eq:E_scat_expansion}) is only valid outside of the smallest sphere circumscribing the scatterer, there is no restriction in the above derivation regarding the surface of integration provided it is outside of the interior volume of the scatterer. It is even possible to apply Eqn.~(\ref{eq:final_with_J3}) on the surface of the scatterer, independently of its shape, as we will later show in an example.

After implementing expression~(\ref{eq:final_with_J3}) into a FEM solver, we observed large errors in higher order multipoles when decomposing the fields caused by small scatterers. We found that the reason for this numerical error are the singularities of $\mathbf{M}_{m,n}^{(3)}$ and $\mathbf{N}_{m,n}^{(3)}$ at the origin of the coordinates. The numerical error produced by the singular fields increases the closer the integration surface is to the origin of the coordinates and the higher the multipole order is. 

Furthermore, these singularities make the decomposition very sensitive to numerical noise, amplifying the noise frequently present in the solutions of the scattered fields computed using numerical solvers.

\begin{figure}[htpb]
\centering
\subfloat{ 
\includegraphics[width=\linewidth]{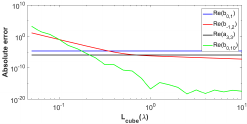}
}
\hfill
\subfloat{
\includegraphics[width=\linewidth]{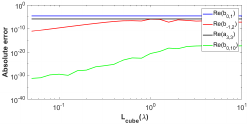}
}
\caption{Top: Decomposition of the scattered field $(0.5+0.5i)\mathbf{M}_{0,1}^{(3)}+0.0277\mathbf{N}_{3,3}^{(3)}$ into the multipoles $\mathbf{M}_{0,1}^{(3)}$, $\mathbf{M}_{-1,2}^{(3)}$, $\mathbf{N}_{3,3}^{(3)}$ and $\mathbf{M}_{0,10}^{(3)}$ using expression~(\ref{eq:final_with_J3}). The integration surface consist of a cube of length $L_\mathrm{cube}$ centered at the origin of coordinates. Bottom: Same decomposition using expressions~(\ref{eq:final_M_decomposition}) and (\ref{eq:final_N_decomposition}).}
\label{fig:orthogonality_J3}
\end{figure}

To analyze and to illustrate this problem, we artificially generated a scattered field composed of the multipoles $\mathbf{M}_{0,1}^{(3)}$ and $\mathbf{N}_{3,3}^{(3)}$ and computed their analytical expansion. We chose the amplitude of the higher order multipole such that there is a strong near field with an amplitude of around 20 V/m at a distance of $\lambda/10$ away from the origin of coordinates. The actual field considered was $\left((0.5+0.5i)\mathbf{M}_{0,1}^{(3)} + 0.0277\mathbf{N}_{3,3}^{(3)}\right)$ V/m. This field was then decomposed using a cubical surface surrounding the origin of coordinates. The implementation of expression~(\ref{eq:final_with_J3}) was done using a trapezoidal integration with a regular grid of 40,000 points for each of the faces of the cube. We then tried to decompose the scattered field into the multipole contributions $\mathbf{M}_{0,1}^{(3)}$, $\mathbf{M}_{-1,2}^{(3)}$, $\mathbf{N}_{3,3}^{(3)}$, $\mathbf{M}_{0,10}^{(3)}$. The error of the decomposition is shown in the top graph of Fig.~\ref{fig:orthogonality_J3} as a function of the length of the cube edge. It can be seen that for the amplitudes absent in the actual source field, the error exponentially increases the closer the integration surface is to the origin of coordinates. Since our final interest is to have an efficient tool for the decomposition of the field obtained from numerical solvers, it is important that the integration surface can be close to the scatterer in order to reduce the computational domain.

To mitigate the numerical stability problems, we need to replace the basis functions $\mathbf{M}_{m,n}^{(3)}$ and $\mathbf{N}_{m,n}^{(3)}$ by some regular fields, which fulfil an orthogonal relation equivalent to expression~(\ref{eq:J3_far_field_identity}). The most natural solution are the regular\cite{mishchenko2002scattering} vector spherical wave functions $\mathbf{M}_{m,n}^{(1)}$ and $\mathbf{N}_{m,n}^{(1)}$. These fields also fulfil Maxwell's equations in homogeneous media but not the Silver-M\"uller radiation condition, i.e. they are frequently used to expand the incident field. In Appendix \ref{sec:Appendix_regular_VSWF} we demonstrate the following relations that allow to calculate the amplitudes of the multipole moments as

\begin{equation}\label{eq:final_M_decomposition}
\begin{split}
a_{m,n} = -ik \int_{\Gamma _1}&{\bigg\{\left(\nabla\times\mathbf{E_{scat}}\right)\times\mathbf{N}_{m,n}^{(1)*}} \\
&-k\mathbf{M}_{m,n}^{(1)*}\times\mathbf{E_{scat}}\bigg\}\cdot\mathbf{dS},
\end{split}
\end{equation}

\begin{equation}\label{eq:final_N_decomposition}
\begin{split}
b_{m,n} = -ik\int_{\Gamma _1}&{\bigg\{\left(\nabla\times\mathbf{E_{scat}}\right)\times\mathbf{M}_{m,n}^{(1)*}} \\
&-k\mathbf{N}_{m,n}^{(1)*}\times\mathbf{E_{scat}}\bigg\}\cdot\mathbf{dS}.
\end{split}
\end{equation}
Note that $a_{m,n}$ and $b_{m,n}$ are the coefficients of the expansion of $\mathbf{E_{scat}}$ into the radiative fields $\mathbf{N}_{m,n}^{(3),*}$ and $\mathbf{M}_{m,n}^{(3),*}$, even though $\mathbf{N}_{m,n}^{(1),*}$ and $\mathbf{M}_{m,n}^{(1),*}$ are used in the expressions. 
We note that equations (\ref{eq:final_M_decomposition}) and (\ref{eq:final_N_decomposition}), when particularized to the boundary of a scatterer, are equivalent to Eqn. 5.175 of Ref.~\onlinecite{mishchenko2002scattering}, derived in the context of the extended boundary condition method for computing the T-matrix of simple homogeneous particles.

Unlike the spherical Hankel functions, the spherical Bessel functions involved into the regular VSWF do have zeros for some values of the spherical coordinate $r$. Note that this fact does not affect the stability of the decomposition, as expressions~(\ref{eq:final_M_decomposition}) and~(\ref{eq:final_N_decomposition}) are functions of both the electric and magnetic field. Therefore, one of the terms of the integrand will always be non-zero in the radial regions where the other term vanishes.

Computing again the same decomposition with the new expression, we got the results shown in the bottom graph of Fig.~\ref{fig:orthogonality_J3}:  the problem with the exponentially increasing error is solved with these new expressions. The remaining existing errors are due to the discretization in the trapezoidal integrations.

\section{Finite Element Method implementation}

We have implemented the derived expressions~(\ref{eq:final_M_decomposition}) and (\ref{eq:final_N_decomposition}) into the commercial finite element solver JCMsuite \cite{JCMsuite_web}. The surface integral is computed on the boundary of the computational domain. 

\begin{figure}[htb]
\subfloat{
\includegraphics[width=\linewidth]{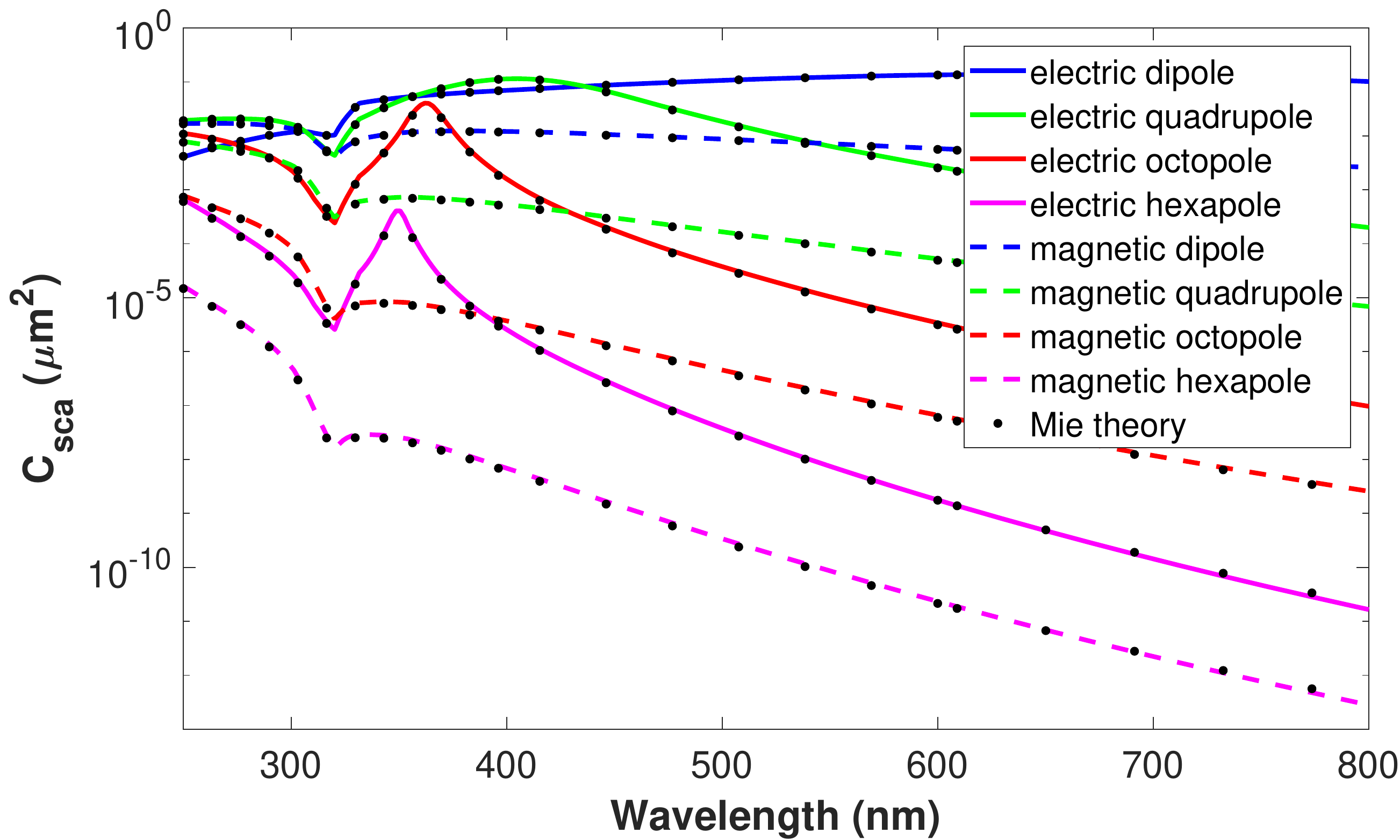}
}
\hfill
\subfloat{
\includegraphics[width=\linewidth]{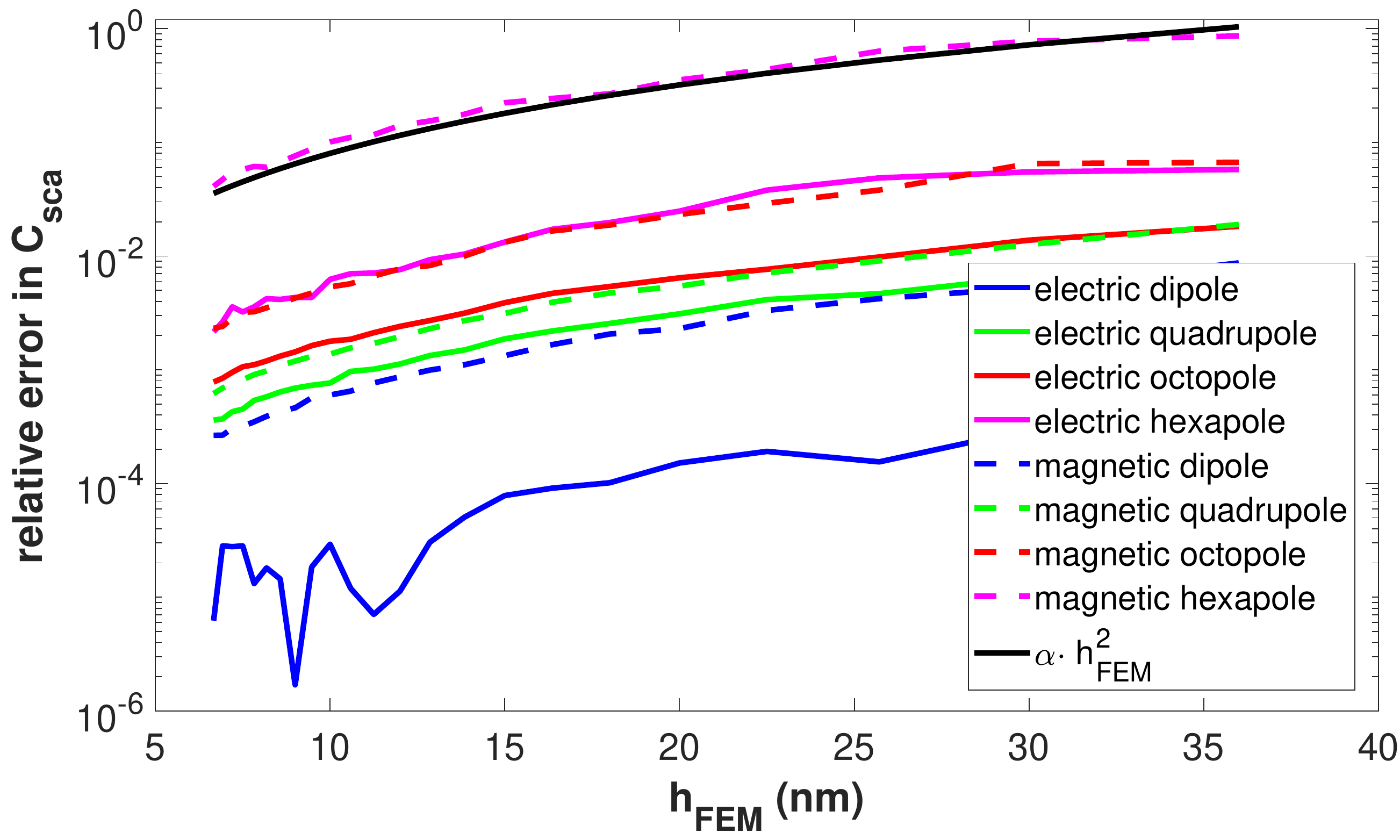}
}
\caption{Top. Spectrally resolved multipole decomposition of the scattering cross section for a silver sphere illuminated with a linearly polarized plane wave. The sphere has a radius of 112.5 nm. The plane wave is propagating in $-z$ direction and the electric field is polarized in $x$ direction. The dots correspond to the results from Mie theory.
Bottom. Convergence study of the multipolar decomposition with respect to the average side length of the tetrahedra FEM discretization, $h_{FEM}$. $h_{FEM}$ indicates the average side length of the mesh elements discretizing the domain surrounding the sphere. The average side length of the tetrahedra in the sphere equals $h_{FEM} / n_{sphere}$, being $n_{sphere}$ the refractive index of the sphere. The study has been done at a wavelength of 450 nm with FEM elements of polynomial order p = 2. The error of the scattering cross section for each multipole order is relative to the value of its cross section.}
\label{fig:FEM_Mie_convergence}
\end{figure}

In order to test the implementation, we calculate the decomposition of the scattered field generated by a sphere when illuminated with a plane wave. The scattered field produced by a homogeneous sphere is a very well known solution and can be computed using the analytical expressions from Mie theory \cite{wriedt_mie}. In the top part of Fig.~\ref{fig:FEM_Mie_convergence} we show the contribution of the different multipoles to the spectral scattering cross section of the sphere. The results were obtained with the FEM implementation (lines) and with Mie theory (dots). In the bottom part of the figure a convergence test of the FEM solution is shown for an illumination wavelength of 450 nm. The errors were calculated with respect to Mie theory. The sphere was placed into the center of a cubic computational domain. The cube has a side length of 1000 nm and the decomposition was performed on its boundary. 

We can see that the results of the decomposition coincide with the analytical results of Mie theory. The validity of the method is also shown in the convergence study. The errors follow a power law with respect to the side length of the mesh elements, being higher for the multipoles with a lower contribution to the total scattered field. We have included in the convergence results the line $\alpha h^p$ for a $\alpha$ value of 8e-4. We can see from this curve that the FEM implementation of the multipole decomposition follows a similar behavior as the asymptotic FEM convergence (cf. Ref.~\onlinecite{FEM_convergence} Section 5.7) for the electromagnetic field values.

\begin{figure}[htb]
\centering
\includegraphics[width=\linewidth]{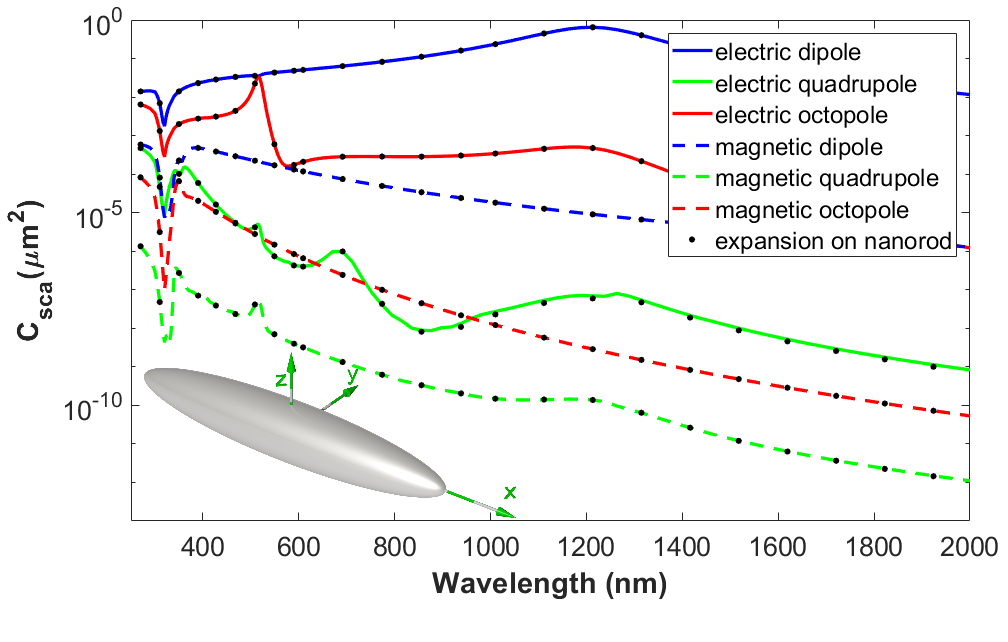}
\caption{Spectrally resolved multipole decomposition of the scattering cross section for a silver nanorod. The nanorod is illuminated with a plane wave propagating in -z direction and a wavelength of 450 nm. The incident electric field is linearly polarized in x. The nanorod is modeled as a body of revolution with an elliptical shape where the major axis is parallel to the $x$ axis. The major and minor semi-axis have a length of 225 and 37.5 nm, respectively. Lines: decomposition performed over the boundary of the computational domain. Dots: decomposition performed over the nanorod surface. } 
\label{fig:FEM_nanorod}
\end{figure}

Finally, we compute the spectral multipole contribution of the scattered field of a silver nanorod illuminated with a linearly polarized plane wave. The nanorod is modelled as a body of revolution with a major and minor semi-axis of 225~nm and 37.5~nm, respectively. The schematic of the simulated structure is shown in the inset of Fig.~\ref{fig:FEM_nanorod}. 

For this specific scatterer, the use of expressions~(\ref{eq:final_M_decomposition})-(\ref{eq:final_N_decomposition}) leads to a reduction in the computational costs of around 70\% with regards to Eqns.~(\ref{eq:a_decomposition})-(\ref{eq:b_decomposition}). This is because the nanorod is four times longer in the $X$ axis than in the $Y$ and $Z$ axis. Therefore, the volume of the computational domain needed to surround the nanorod can be approximately four times smaller than if it had to contain a sphere surrounding the scatterer. To be precise, the computational domain consisted of a cuboid with side lengths of 562.5 nm, 300 nm, and 300 nm respectively. The incident plane wave is propagating in the $-z$ direction, with an amplitude of 1 V/m and is polarized in $x$ direction. The results are shown in Fig.~\ref{fig:FEM_nanorod}. To show that the decomposition is valid over any surface surrounding the scatterer, we performed the decomposition over the computational domain, represented by lines, and also on the surface of the scatterer, represented by dots. It can be clearly seen, that the results are indistinguishable.

\begin{figure}[htb]
\includegraphics[width=\linewidth]{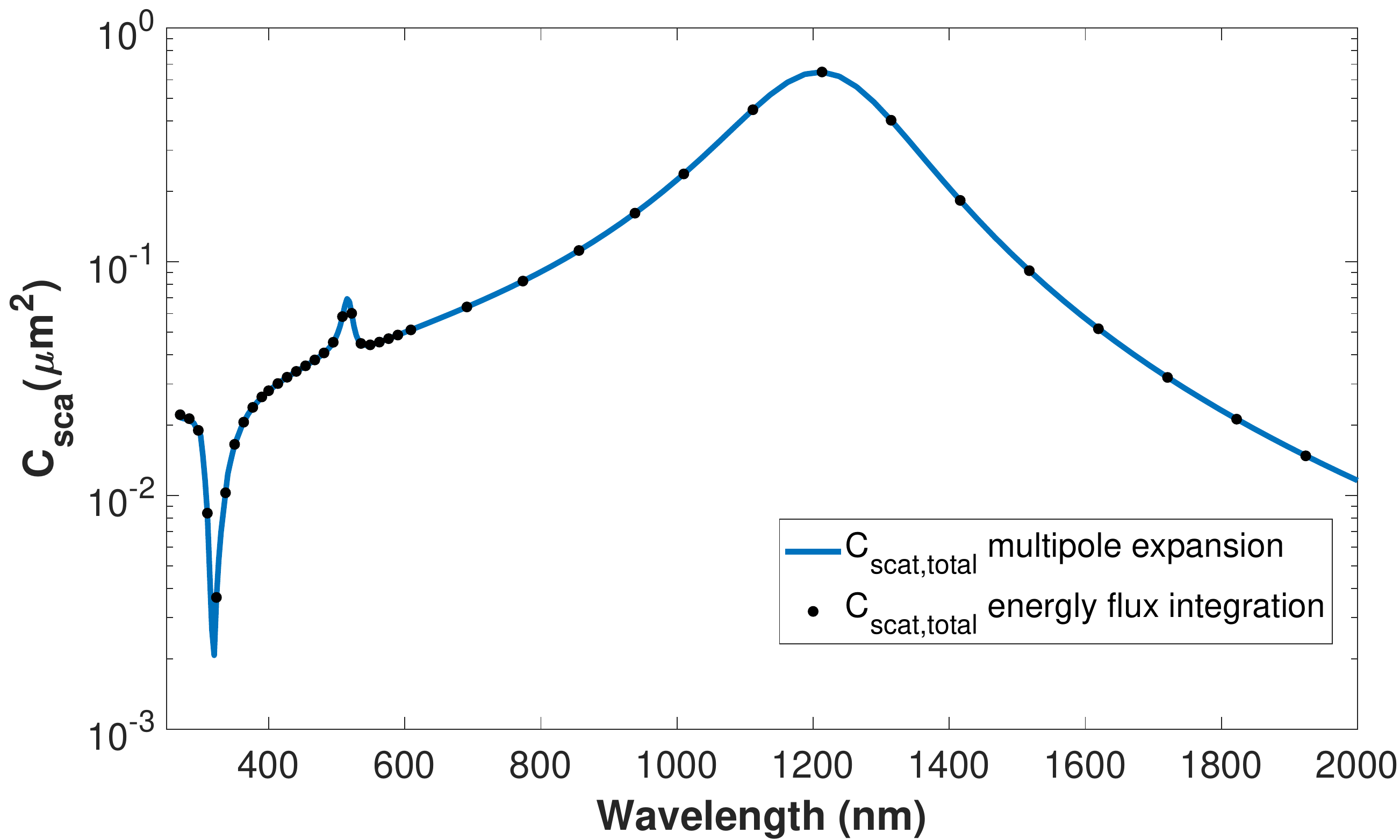}
\caption{Comparison between the values of the total scattering cross section obtained based on the integral of the scattering energy flux and based on the results of the multipole expansion.}
\label{fig:comparison_scat_total}
\end{figure}

We can see how the resonance shifted to longer wavelengths compared with the sphere and how the scattered field is mainly an electric dipolar field at these frequencies. In Fig.~\ref{fig:comparison_scat_total} we compare the total scattered cross section calculated based on the results of the multipole expansion and also with the calculation of the integral of the time averaged scattered energy flux $<\mathbf{S}>$

\begin{equation}
C_{scat,total} = \frac{\int_{S}{\left<\mathbf{S}\right> \cdot \mathbf{dS}}}{2Z^2|E_i|^2},
\end{equation}
where $Z$ the characteristic impedance of the surrounding medium and $E_i$ the amplitude of the incident plane wave.

We can see that the results obtained from both methods coincide. The total scattering cross section at the resonance wavelength is 0.648 $\mathrm{\mu m^2}$. For the sphere, the resonance peak of the total scattering cross section has a value of 0.27 $\mathrm{\mu m^2}$ at a wavelength of 364 nm and it has 4 mainly multipole contributions. The geometrical cross sections are 0.027 $\mathrm{\mu m^2}$ and 0.039 $\mathrm{\mu m^2}$ for the nanorod and the sphere respectively.

\section{Conclusions}

We have proposed an orthonormal relation in order to implement the multipole expansion of scattered fields by integrals on generic surfaces. This is particularly suited for use in combination with numerical solvers relying on methods such as FDTD or FEM. The main advantages of this implementation are the convenience and a better tradeoff between accuracy and computational costs, as no interpolations of the scattered field needs to be computed. We have also shown how to use the regular VSWF in order to decompose the scattered field into the radiative VSWF. The use of regular fields significantly improves the numerical stability of the implementation. Finally, we have tested the implementation using Mie theory and shown one example for a structure which can not be decomposed with analytical means.

\appendix

\section{Radiative and regular VSWF}\label{sec:Appendix_VSWF}

In this work we use the below definition for the vector spherical wave functions $\mathbf{M_{m,n}^{J}}(\mathbf{r})$ and $\mathbf{N_{m,n}^{J}}(\mathbf{r})$ in spherical coordinates,

\begin{equation}
\mathbf{M_{m,n}^{J}}(\mathbf{r}) = E_{mn}\left(0, i\pi _{mn}\left(\theta\right), -\tau _{mn}\left(\theta\right)\right)Z_n^J\left(kr \right)e^{im\phi},
\end{equation}

\begin{equation}
\mathbf{N_{m,n}^{J}}(\mathbf{r}) = E_{mn}\left(N_{m,nr}^{J},N_{m,n\theta}^{J},N_{m,n\phi}^{J} \right)e^{im\phi}
\end{equation}

\begin{align}
N_{m,nr}^{J}(\mathbf{r}) &= n\left(n+1\right)P_{nm}(\cos{\theta})\frac{Z_n^J(kr)}{kr}, \\
N_{m,n\theta}^{J}(\mathbf{r}) &= \tau_{mn}\left(\theta\right) \frac{-nZ_n^J(kr)+krZ_{n-1}^J\left(kr\right)}{kr}, \\
N_{m,n\phi}^{J}(\mathbf{r}) &=  i\pi_{mn}\left(\theta\right)\frac{-nZ_n^J\left(kr\right)+krZ_{n-1}^J\left(kr\right)}{kr},
\end{align}
with $E_{mn}$ being the normalization factor given by Eqn.~C.25 of Ref.~\onlinecite{mishchenko2002scattering}

\begin{equation}\label{eq:normalization}
E_{mn} = \sqrt{\frac{\left(2n+1\right)\left(n-m\right)!}{4\pi\left(n+1\right)\left(n+m\right)!}}.
\end{equation}

$\pi _{mn}$ and $\tau _{mn}$ functions are defined in terms of the associated Legendre polynomials,

\begin{align}
\pi _{mn} (\theta) &= \frac{m}{\sin{\theta}}P_{nm}(\cos{\theta}) \\
\tau _{mn} (\theta) &= \frac{n}{\tan{\theta}}P_{nm}(\cos{\theta}) -\frac{\left(n+m\right)}{\sin{\theta}}P_{n-1,m}(\cos{\theta}), \\
P_{nm} (\theta) &= (-1)^m P_n^m = \frac{\left(1-x^2\right)^{m/2}}{2^ll!}\frac{d^{l+m}}{dx^{l+m}}\left(x^2-1\right)^l.
\end{align}

The superscript $J$ indicates the radial dependence of the VSWFs in terms of the Bessel functions $J_{n+0.5}(x)$ and $Y_{n+0.5}(x)$,

\begin{align}
Z_n^{(1)}(x) &= j_n(x) = \sqrt{\frac{\pi}{2x}}J_{n+0.5}(x), \\
Z_n^{(2)}(x) &= y_n(x) = \sqrt{\frac{\pi}{2x}}Y_{n+0.5}(x), \\
Z_n^{(3)}(x) &= j_n(x) +iy_n(x), \\
Z_n^{(4)}(x) &= j_n(x) -iy_n(x). 
\end{align}

With the normalization factor given by Eqn.~(\ref{eq:normalization}), the power radiated by a scattered field decomposed into VSWF as given in Eqn.~(\ref{eq:E_scat_expansion}) can be calculated as

\begin{equation}
P = \frac{1}{2Zk^2}\sum_{n = 1}^{\infty}{\sum_{m = -n}^{n}{\left(|a_{m,n}|^2+|b_{m,n}|^2\right)}},
\end{equation}
being Z the characteristic impedance of the medium.

\section{Replacing radiative VSWF by regular VSWF}\label{sec:Appendix_regular_VSWF}

In this appendix, we show that substituting the radiative VSWF $\left\{\mathbf{M},\mathbf{N}\right\}_{m,n}^{(3)}$ by the regular VSWF $\left\{\mathbf{M},\mathbf{N}\right\}_{m,n}^{(1)}$ as test functions in expression~(\ref{eq:surface_integrals_F}) ends in a relation proportional to Eqn.~(\ref{eq:J3_far_field_identity}). This new relation lets us express the decompositions~(\ref{eq:a_decomposition}) and (\ref{eq:b_decomposition}) as an integral over a general surface involving the scattered field and the regular VSWF.
In order to simplify the expressions, in this section we will just use now the VSWF $\mathbf{M}_{m,n}^{(1)}$. The same procedure can be applied to the fields $\mathbf{N}_{m,n}^{(1)}$ giving equivalent results.

Let's start from Eqn.~(\ref{eq:I2}) by substituting $\mathbf{F}$ with $\mathbf{M}_{m,n}^{(1)}$,

\begin{align}
I_2 = \int_{S^2_R}&{\left\{\mathbf{dS}\times\left(\nabla\times\mathbf{E_{scat}}\right)\right\}\cdot\mathbf{M}_{m,n}^{(1)*}}\nonumber \\
                  -&\left\{\mathbf{dS}\times\left(\nabla\times\mathbf{M}_{m,n}^{(1)*}\right)\right\}\cdot\mathbf{E_{scat}}. \label{eq:I2_J1}
\end{align}
The regular VSWF $\mathbf{M}_{m,n}^{(1)}$ can be decomposed into a combination of radiative $\mathbf{M}_{m,n}^{(3)}$ and $\mathbf{M}_{m,n}^{(4)}$ vector spherical wave functions, fulfilling an outwards and inwards radiation condition, respectively,

\begin{equation}\label{eq:decomposition_J1}
\mathbf{M}_{m,n}^{(1)} = \frac{\mathbf{M}_{m,n}^{(3)} + \mathbf{M}_{m,n}^{(4)}}{2},
\end{equation}
\begin{equation}\label{eq:Silver-Mueller_J3}
\lim_{R\to\infty}{\left(\nabla\times\mathbf{M}_{m,n}^{(3)}\right)\times R\mathbf{\hat{r}}} =  +\lim_{R\to\infty}{ikR\mathbf{M}_{m,n}^{(3)}},
\end{equation}
\begin{equation}\label{eq:Silver-Mueller_J4}
\lim_{R\to\infty}{\left(\nabla\times\mathbf{M}_{m,n}^{(4)}\right)\times R\mathbf{\hat{r}}} =  -\lim_{R\to\infty}{ikR\mathbf{M}_{m,n}^{(4)}}.
\end{equation}
By plugging Eqn.~(\ref{eq:decomposition_J1}) into Eqn.~(\ref{eq:I2_J1}) we get,

\begin{align}
I_2 = \int_{S^2_R}&{\left\{\mathbf{dS}\times\left(\nabla\times\mathbf{E_{scat}}\right)\right\}\cdot\frac{\mathbf{M}_{m,n}^{(3)*}+\mathbf{M}_{m,n}^{(4)*}}{2}} \nonumber \\
                 -&\left\{\mathbf{dS}\times\left(\nabla\times\frac{\mathbf{M}_{m,n}^{(3)*}+\mathbf{M}_{m,n}^{(4)*}}{2}\right)\right\}\cdot\mathbf{E_{scat}}.
\end{align}
By applying now Silver-M\"uller radiation conditions (\ref{eq:Silver-Mueller_J3})-(\ref{eq:Silver-Mueller_J4}), we get
\begin{align}
\lim_{R\to\infty} I_2 &= \lim_{R\to\infty} -ik\int_{S^2_R}{\Bigg(\mathbf{E_{scat}}\cdot\frac{\mathbf{M}_{m,n}^{(3)*}+\mathbf{M}_{m,n}^{(4)*}}{2}} \nonumber \\
                      &+ \frac{\mathbf{M}_{m,n}^{(3)*}-\mathbf{M}_{m,n}^{(4)*}}{2}\cdot\mathbf{E_{scat}}\Bigg)\,\mathrm{dS} \nonumber \\
&= \lim_{R\to\infty} -ik\int_{S^2_R}{\mathbf{E_{scat}}\cdot\mathbf{M}_{m,n}^{(3)*} \,\mathrm{dS}}. \label{eq:final_I2_M1}
\end{align}
Finally, using Eqn.~(\ref{eq:surface_integrals_F})

\begin{align}
&\int_{\Gamma _1}{\bigg\{\left(\nabla\times\mathbf{E_{scat}}\right)\times\mathbf{M}_{m,n}^{(1)*} }
-\left(\nabla\times\mathbf{M}_{m,n}^{(1)*}\right)\times\mathbf{E_{scat}}\bigg\}\cdot\mathbf{dS} \nonumber \\
&= \lim_{R\to\infty} ik\int_{S^2_R}{\mathbf{E_{scat}}\cdot\mathbf{M}_{m,n}^{(3)*}\,\mathrm{dS}}
\end{align}
and using the fact that $\nabla\times\mathbf{M}_{m,n}^{(1)}$ equals $k\mathbf{N}_{m,n}^{(1)}$, we obtain

\begin{align}
&\int_{\Gamma _1}{\bigg\{\left(\nabla\times\mathbf{E_{scat}}\right)\times\mathbf{M}_{m,n}^{(1)*} }
-k\mathbf{N}_{m,n}^{(1)*}\times\mathbf{E_{scat}}\bigg\}\cdot\mathbf{dS} \nonumber \\
&= \lim_{R\to\infty} ik\int_{S^2_R}{\mathbf{E_{scat}}\cdot\mathbf{M}_{m,n}^{(3)*}\,\mathrm{dS}}. \label{eq:M1_far_field_identity}
\end{align}

An analogous procedure gives an equivalent relation for fields $\mathbf{N}_{m,n}^{(1)}$,

\begin{align}
&\int_{\Gamma _1}{\bigg\{\left(\nabla\times\mathbf{E_{scat}}\right)\times\mathbf{N}_{m,n}^{(1)*} }
-k\mathbf{M}_{m,n}^{(1)*}\times\mathbf{E_{scat}}\bigg\}\cdot\mathbf{dS} \nonumber \\
&=\lim_{R\to\infty} ik\int_{S^2_R}{\mathbf{E_{scat}}\cdot\mathbf{N}_{m,n}^{(3)*}\,\mathrm{dS}}. \label{eq:N1_far_field_identity}
\end{align}

These expressions are just the desired relations corresponding to Eqn.~(\ref{eq:J3_far_field_identity}) but with the regular VSWF. They can be used analogously to derive Eqns.~(\ref{eq:final_M_decomposition}) and (\ref{eq:final_N_decomposition}) of the main manuscript.

\begin{acknowledgments}
We thank Philipp Gutsche for his work on the VSWF code and for his contribution to test the implementation.\\
This project has received funding from the European Union's Horizon 2020 research and innovation programme under the Marie Sklodowska-Curie grant agreement No 675745. \\
This project has received funding from Deutsche Forschungsgemeinschaft (DFG) grant RO 3640/7-1. \\
We acknowledge the support from the Karlsruhe School of Optics and Photonics (KSOP). \\
This work is partially funded through the project 17FUN01 "BeCOMe" within the programme EMPIR. The EMPIR initiative is co-founded by the European Union's Horizon 2020 research and innovation programme and the EMPIR Participating Countries.\\
\end{acknowledgments}


%

\end{document}